# On the Impact of the Model-Based Representation of Inconsistencies to Manual Reviews

## Results from a Controlled Experiment – Extended Version[1]

Marian Daun, Jennifer Brings, Thorsten Weyer

paluno – The Ruhr Institute for Software Technology
University of Duisburg-Essen, Essen, Germany

{marian.daun, jennifer.brings,
thorsten.weyer}@paluno.uni-due.de

**Abstract.** To ensure fulfilling stakeholder wishes, it is crucial to validate the documented requirements. This is often complicated by the fact that the wishes and intentions of different stakeholders are somewhat contradictory, which manifests itself in inconsistent requirements. To aid requirements engineers in identifying and resolving inconsistent requirements, we investigated the usefulness for manual reviews of two different model-based representation formats for inconsistent requirements; one that represent the inconsistent requirements in separate diagrams and one that represents them integrated into one diagram using annotations. The results from a controlled experiment show that the use of such integrated review diagrams can significantly increase efficiency of manual reviews, without sacrificing effectiveness.

## 1      Motivation and Background

Model-based engineering has widely been adopted in the domain of embedded systems to cope with the growing complexity of such systems [1–3]. Model-based requirements engineering is often seen as an important part as it allows, among others, for full continuity across the entire engineering process [4]. As model-based documentation is often used from different requirements perspectives or to document the intentions of different stakeholders in different requirements models [5], inconsistencies between multiple requirements models can easily arise and need to be investigated and resolved by requirements engineers [6]. This particular challenge can often only be solved by manual validation, as automated approaches can only detect inconsistencies but not negotiate agreement between different stakeholders and thus, ensure the correctness of all requirements artifacts (cf. [7]). Manual validation is needed to ensure that the "documentation […] meet[s] user needs and expectations, whether specified or not" [8]. Additionally, for safety critical systems such as those embedded in vehicles or airplanes for example, safety standards (e.g., ISO 26262 [9] concerning the functional safety of road vehicles, ARP 4761 [10] concerning the safety assessment of civil aircraft) mandate the use of manual validation techniques.

---

[1] This is an extended version of *Daun M., Brings J., Weyer T. (2017) On the Impact of the Model-Based Representation of Inconsistencies to Manual Reviews. In: Mayr H., Guizzardi G., Ma H., Pastor O. (eds) Conceptual Modeling. ER 2017. Lecture Notes in Computer Science, vol 10650. Springer, Cham* https://doi.org/10.1007/978-3-319-69904-2_35 containing additional experiment materials.

Message sequence charts like languages (e.g., ITU Message Sequence Charts [11], UML Sequence Diagrams [12], Life Sequence Charts [13]) are commonly used in requirements engineering, to represent typical system execution traces in scenarios (e.g., [14]) or to define the intended interaction-based behavior of embedded systems (e.g., [15]). While message sequence charts have shown to be an effective and efficient language to be used in manual reviews (cf. [16]), the validation of different inconsistent requirements models is challenging because it involves investigating a potentially vast number of diagrams containing identical parts, alternative parts, and contradictory parts. To reduce the number of diagrams to be reviewed, existing automated model merging techniques (e.g., [14]) can create diagrams that represent inconsistent properties integrated in just one diagram.

This paper reports on a controlled experiment to investigate whether model merging can be used to improve effectiveness and efficiency of reviews. Therefore, the experiment compares two representations of inconsistencies in ITU Message Sequence Chart models: the first shows inconsistent properties in two separate diagrams, and the second showing just one integrated diagram, highlighting the inconsistencies. Results show that while the representation of inconsistencies has no significant impact on the review's effectiveness; for diagrams with few inconsistencies, it is significantly more efficient to use the integrated representation.

To introduce the experiment, Section 2 gives an overview of related approaches dealing with manual reviews of model-based specifications as well as related experiment reports, which were used as basis for the definition of the experiment design. Subsequently, Section 3 introduces the notation of ITU Message Sequence Charts and the concrete representation formats for the separate representation of inconsistencies and integrated representation. Section 4 defines the experimental setup and Section 5 reports on the results of the experiment. Finally, Section 6 concludes the paper with a discussion of the major findings, threats to validity and inferences.

## 2   Related Work

### 2.1   Manual Validation of Model-based Specifications

Perspective-based review (also referred to as perspective-based reading) is a widely accepted inspection technique for requirements specifications (cf.[17]). In general, the requirements are validated from the perspective of a certain later development phase (e.g., from the perspective of a designer, a tester, or a safety engineer). Thereby, the review aims at the early detection of different kinds of defects. In [18], *Denger and Ciolkowski* describe a defect taxonomy to apply a perspective-based inspection technique to statecharts. In addition, *Binder* defines a checklist to validate statecharts from a testing perspective in [19]. A more general approach to validate model-based specifications is presented by *Travassos et al.* in [20]. The approach addresses the consistency between UML diagrams of different types. For this purpose, perspective-based reviews of scenarios from different perspectives are suggested.

In previous work, we proposed the use of dedicated review models to support the validation of embedded systems functional design against stakeholder intentions [21] and to support the validation of the behavioral requirements [22]. Therefore, we investigated the industrial feasibility of review models [23] and the effects of the used modeling language for conducting reviews from a requirements perspective finding that the use of ITU message sequence charts as review artifact is more effective, efficient, and subjectively supportive compared to the use of the original specification of the embedded systems' functional design (cf. [16]).

### 2.2   Related Experiments

Several studies already investigated effectiveness and efficiency of different review types. Therefore, the experiment design presented in this paper is based on the experiment design of these reported experiments. Most of these studies deal with comparing the effectiveness of perspective-based reviews against checklist-based reviews, which is often seen as standard inspection technique. Based on common defects a checklist is created to manually validate whether this defect is included in the specification.

For example, *Miller et al.* [24] report on a student experiment with 50 trained students, finding that perspective-based reviews are more effective than checklist-based approaches for error detection in natural language requirements specifications. *Basili et al. 1996* [25] report on a controlled experiment with professional software developers, where the perspective-based review is evaluated as significantly more effective than other inspection techniques for requirements documents. In replications conducted by *He and Carver* [26] as well as *Maldonado et al.* [27] this finding is supported. Also *Porter et al.* [28] and [29]*, Laitenberger et al.* [30]*, Berling and Runeson* [31]*, and Sabaliauskaite et al.* [32] report similar experiments with comparable findings.

## 3 Foundations

Message sequence charts describe the system's behavior in terms of execution paths between the system and its environment. Message sequence charts are standardized by ITU recommendation Z.120 [11], which defines a graphical and a textual notation. Furthermore, several formalizations in accordance with Z.120 are given (e.g., [33]), which allow for the application of formal model transformation and merging techniques. **Fig. 1** shows the graphical notation of a basic Message Sequence Chart (bMSC). Beside bMSCs, Z.120 defines high-level Message Sequence Charts (hMSCs) to give the interaction paths defined by several single bMSC a global order.

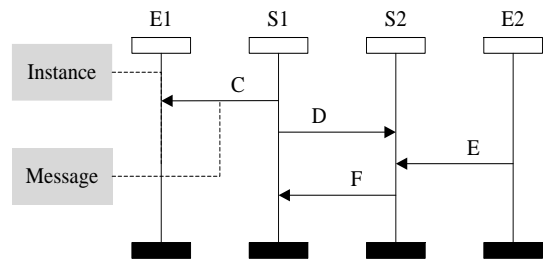

**Fig. 1.** Graphical notation of a basic message sequence chart

For various reasons (e.g., for view merging [34], for comparing different orderings of scenarios [35] or slicing of message sequence charts specifications [33]) model transformations with message sequence charts like languages as input (e.g., [36]) or output (e.g., [37]) have been defined as well as merging algorithms to consolidate different diagrams (e.g., [14]). In common merging approaches for ITU Message Sequence Charts, the bMSCs are not merged directly, but after transformation into automaton notation. This seems unsurprising since Z.120 defines automata semantics for the execution of bMSCs and merging approaches for finite automata are well-established. This is, moreover, advantageous as it allows for merging of bMSCs under consideration of the structure given by the hMSC and, hence, allows for comparison of different structures and merging of Message Sequence Charts specifications which have been sliced differently.

**Fig. 2** illustrates a very simple merge of just two separate basic message sequence charts (bMSCs) ((a1) and (a2) in **Fig. 2**). Both bMSCs show the same excerpt of a specification of an automotive lane keeping support system (LKS). Diagram (a1) shows how the system shall handle lane departures from the perspective of one stakeholder and diagram (a2) shows how this functionality is specified from another stakeholder`s perspective. As can be seen, these two diagrams differ in just one message that the LKS either receives the current steering angle from the electronic stability support (ESS) or not.

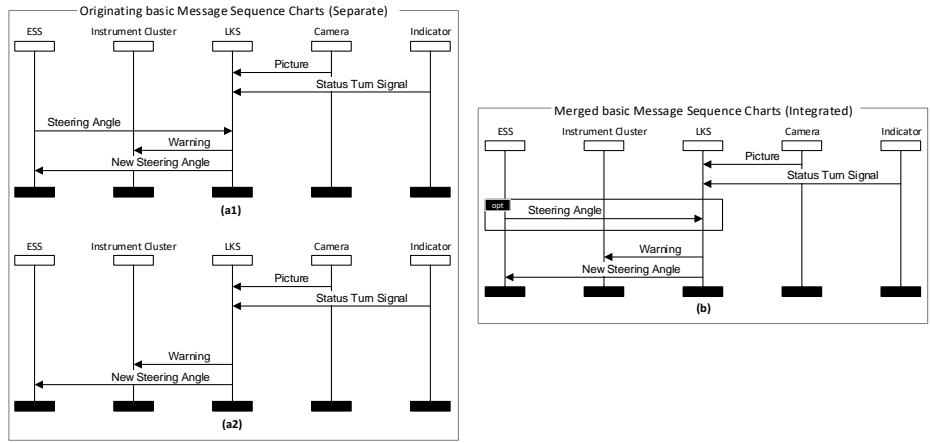

**Fig. 2.** Exemplary model merging of bMSCs

Diagram (b) shows the merged behavior in a corresponding bMSC, which displays both interaction sequences specified by the originating bMSC. In the following, we will refer to the original bMSCs as the separate representation and the merged bMSC as the integrated representation.

## 4 Experiment Planning

While Section 3 introduced the approach under investigation, i.e. the use of one integrated MSC to review inconsistent behavioral properties in contrast to the use of two separate bMSCs, this section introduces the experiment setup to evaluate this approach. To ensure comparability with other controlled experiments the structure of the subsections is based on the recommendations given in [38].

### 4.1 Goals

The experiment shall contribute to the overall research question:

*Is it beneficial for manual reviews to first merge inconsistent behavioral properties into one integrated diagram compared to the review of inconsistent properties in separate diagrams?*

Hence, the goal of our study is to investigate, whether the integrated representation (i.e. the merged diagram) is advantageous compared to the separate representation (i.e. the original bMSC diagrams) for reviews with respect to their *effectiveness*, and *efficiency*.

### 4.2 Participants

As experiment participants students were chosen as recent research has shown, that the use of student participants better allows deriving significant conclusion and must be seen as a fairer experiment between treatment and control technique then the use of industry professionals [39, 40]. However, experimental setup and experiment results were discussed with industry partners to ensure generalizability.

The experiment was conducted with graduate students within a master-level university course on requirements engineering. The experiment was conducted after lessons on validation techniques for requirements. The participants are mainly holding bachelor degrees in 'Systems Engineering' (with particular emphasis on software engineering) or 'Business Information Systems' and are now enrolled in the respective master degree programs. The experiment was conducted with 41 graduate students.

The students were recruited by a mandatory exercise within the course. No bonuses with regard to the courses final exam were given to avoid social threats to validity. In so far, the recruitment strategy has to be seen as opportunity sampling. We used a within-subject design; all participants acted as treatment and control. The order in which the participants were assigned to treatment and to control exercises was randomized.

### 4.3 Experiment Material

We used an industrial sample specification from the avionics domain as source for the experiment material. The participants had to conduct a review of specification artifacts from an avionics collision avoidance system. Additionally, the experiment material contains a set of natural language stakeholder intentions for the avionics collision avoidance system. To ensure generalizability to a real industrial engineering process, we prepared the experiment material in close collaboration with industrial professionals from a large European company in the avionics domain. **Fig. 3** shows an excerpt from the experimental material used.

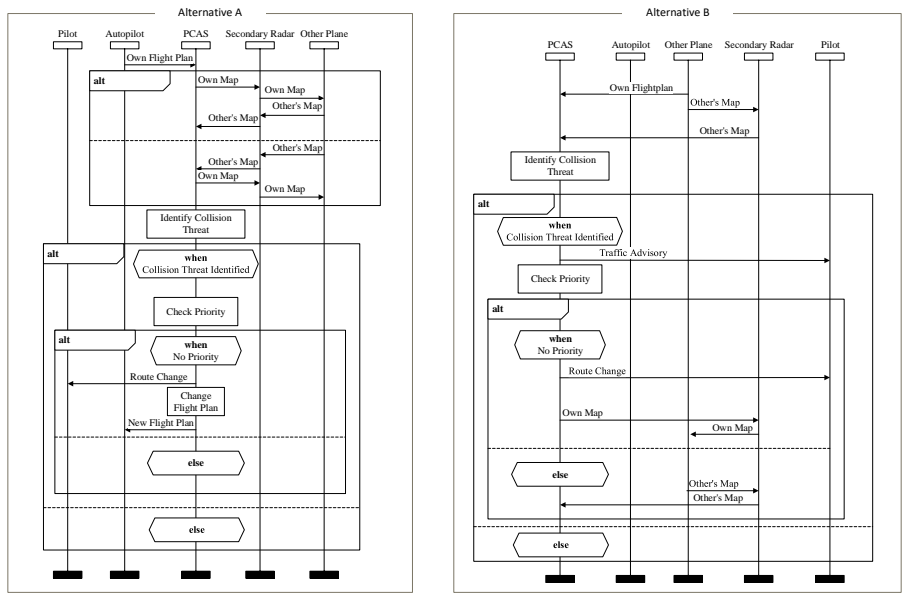

**Fig. 3.** Excerpt from the Experiment Material

### 4.4 Hypotheses and Variables

As independent variables, we investigated the representation format of the review artifact as well as the degree of consistency. First, we differentiated between two different representation formats to validate the specified behavior against the actual stakeholder intentions and second, we differentiated between two degrees of consistency between the artifacts to be validated:

**Table 1.** Independent variables

| Representation Format | Degree of Consistency |
|---|---|
| *Integrated (short: I):* The participants reviewed a diagram using the integrated representation, which displays inconsistencies in the same bMSC. | *High (short: H):* The participants reviewed diagrams in *I* or *S*, which are highly consistent, i.e., that have only few inconsistencies between each other. |
| *Separate (short: S):* The participants reviewed diagrams using the separate representations which displays inconsistencies in two separate bMSCs. | *Low (short: L):* The participants reviewed diagrams in *I* or *S*, which are highly inconsistent, i.e., have many inconsistencies between each other. |

Our distinction between *H* and *L* is based on Miller's law [41], as is commonly done for similar problems. Hence, we introduced less than five inconsistent elements into the diagrams to create diagrams with a high degree of consistency and more than nine to create diagrams with a low degree of consistency. The diagrams themselves were of common sizes in industrial practice, since the examples were excerpts taken from industry specifications. This initial distinction was backed by findings we gained from expert discussions.

As dependent variables, we determined:

- *Effectiveness:* the ratio of correct review decisions made.
- *Efficiency:* the average time spent per correct review decision.

Based on the goals, the independent, and the dependent variables, the null and alternative hypotheses of the experiment can be formulated:

**Table 2.** Hypotheses

| Effectiveness | Efficiency |
|---|---|
| $H_{Eff}$1-0: The review is equally effective no matter the representation. | $H_{Efy}$1-0: The review is equally efficient no matter the representation. |
| $H_{Eff}$1-1a: The review is more effective in *I*. | $H_{Efy}$1-1a: The review is more efficient in *I*. |
| $H_{Eff}$1-1b: The review is more effective in *S*. | $H_{Efy}$1-1b: The review is more efficient in *S*. |
| $H_{Eff}$2-0: The review is equally effective no matter the degrees of consistency. | $H_{Efy}$2-0: The review is equally efficient no matter the degrees of consistency. |
| $H_{Eff}$2-1a: The review is more effective for *H*. | $H_{Efy}$2-1a: The review is more efficient for *H*. |
| $H_{Eff}$2-1b: The review is more effective for *L*. | $H_{Efy}$2-1b: The review is more efficient for *L*. |
| $H_{Eff}$3-0: There is no interaction effect between the representation and the degree of consistency in terms of effectiveness. | $H_{Efy}$3-0: There is no interaction effect between the representation and the degree of consistency in terms of efficiency. |
| $H_{Eff}$3-1: NOT $H_{Eff}$3-0 | $H_{Efy}$3-1: NOT $H_{Efy}$3-0 |

To evaluate the influence of other factors we measured several covariates such as highest educational achievement, degree program, semester, age, as well as participants' self-rated experience in six categories related to conducting reviews in general and the familiarity with Message Sequence Charts.

### 4.5 Experiment Design and Procedure

The experiment used a within-subject design. Each participant conducted a review of an excerpt from the specifications of the avionics collision avoidance system in both representations (*I* and *S*). The order of the reviews using the different representations was randomized for each participant. Each participant was assigned four reviewing tasks (one for each combination of representation format and degree of consistency) and a total of 34 stakeholder intentions for all four tasks. In this setup "reviewing" means to decide for each stakeholder intention which of the following cases applies: 1) the intention is correctly represented in perspective A; 2) the intention is correctly represented in perspective B; 3) the intention is correctly represented in both perspectives; 4) the intention is correctly represented neither in perspective A nor in perspective B. After the review, each participant was asked for some personal data.

The study was conducted as an experiment using an online questionnaire. The experiment was designed to last about 30 minutes to minimize participants' mortality because of losing interest. This setup was chosen on the basis of best practices to mitigate common threats to validity, particularly arising when using student participants [42].

## 5 Analysis

### 5.1 Data Set Preparation

We filtered some of the participants' data sets from the final data set. This was necessary, since some of the participants did obviously not perform serious reviews, since these participants finished the review of all four tasks (consisting of the validation of 34 natural language stakeholder intentions) in less than five minutes. In total, we used 36 data sets for further analysis. We used SPSS 24 to analyze the datasets.

### 5.2 Descriptive Statistics

This section presents the descriptive statistics for effectiveness and efficiency.

*Effectiveness.* Table 3 shows mean, median, and standard deviation for effectiveness for the different representations overall and for the different representations considering the degrees of consistency. When reviewing diagrams with a high degree of consistency, participants made the correct decision when using the integrated representation 81.94% of the time and when using the separate representation 80.56% of the time. When reviewing diagrams with a low degree of consistency, participants made the correct decision when using the integrated representation only 67.93% of the time and when using the separate representation 69.19% of the time. The overall effectiveness is thus 74.94% for the integrated representation and 74.87% for the separate representation.

**Table 3.** Measurements for effectiveness

|     | N Valid | N Missing | Mean | Median | Std. Deviation |
| --- | --- | --- | --- | --- | --- |
| I H | 36 | 0 | 81.94% | 83.33% | 25.63% |
| S H | 36 | 0 | 80.56% | 83.33% | 27.17% |
| I L | 36 | 0 | 67.93% | 72.73% | 20.56% |
| S L | 36 | 0 | 69.19% | 72.73% | 20.42% |
| I   | 36 | 0 | 74.94% | 81.82% | 21.32% |
| S   | 36 | 0 | 74.87% | 81.82% | 22.60% |

**Fig. 4** illustrates the relation between the integrated representation (*I*) and the separate representation (*S*) w.r.t. effectiveness. Many participants made the same number of correct review decisions when using the integrated representation as when using the separate representation ($\varrho(36) = 0.854$, $p < .001$). Note that when reviewing diagrams with a high degree of consistency (*H*), some participants were considerably more effective when using the integrated representation (*I*) than when using the separate representation (*S*). ($\varrho(36) = 0.747$, $p < .001$). However, when reviewing diagrams with a low degree of consistency (*L*) all participants were almost equally effective no matter the representation.

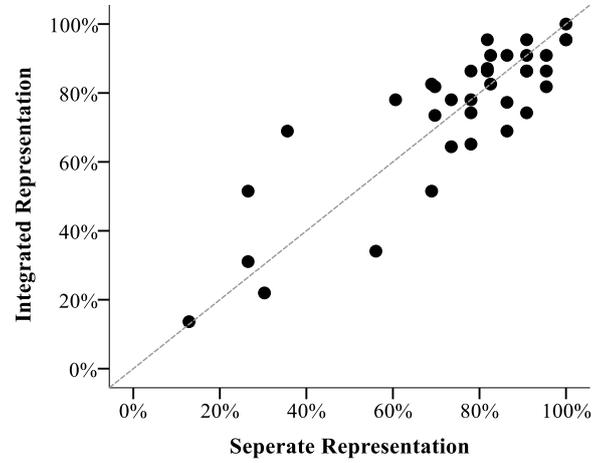

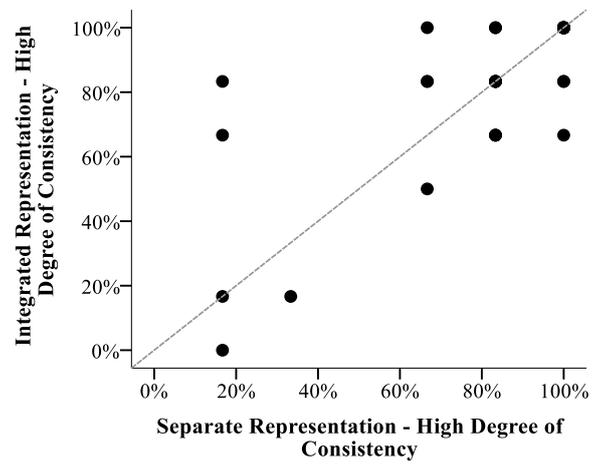

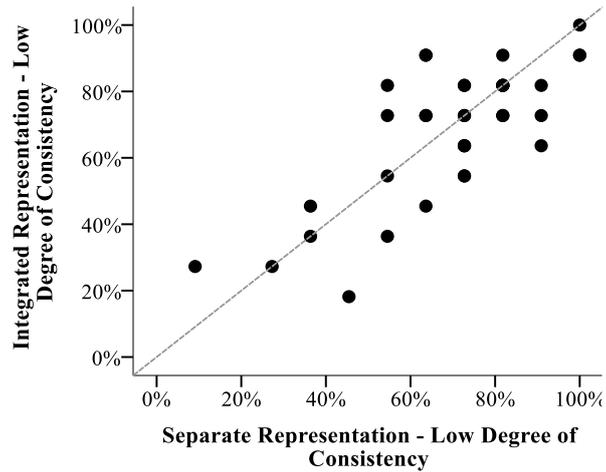

**Fig. 4.** Scatterplots – Effectiveness

*Efficiency.* It took participants on average 52 seconds for each correct decision when using the integrated representation (*I*) and 1.09 minutes when using the separate representation (*S*). When reviewing diagrams with a high degree of consistency (*H*) the average time spent for making a correct decision was 38.4 seconds when using the integrated representation (*I*) and 1.13 minutes when using the separate representation (*S*). When reviewing diagrams with a low degree of consistency (*L*), participants needed on average 1.12 minutes per correct decision using the integrated representation (*I*) and 1.04 minutes using the separate representation (*S*). See Table 4 for further details.

**Table 4.** Measurements for efficiency in minutes

|     | N     |         |      |        |                |
| --- | ----- | ------- | ---- | ------ | -------------- |
|     | Valid | Missing | Mean | Median | Std. Deviation |
| I H | 35    | 1       | 0.64 | 0.56   | 0.39           |
| S H | 36    | 0       | 1.13 | 0.89   | 0.87           |
| I L | 36    | 0       | 1.12 | 0.94   | 0.58           |
| S L | 36    | 0       | 1.04 | 0.90   | 0.60           |
| I   | 36    | 0       | 0.88 | 0.79   | 0.41           |
| S   | 36    | 0       | 1.09 | 0.92   | 0.58           |

**Fig. 5** illustrates the relation between *I* and *S* w.r.t. efficiency. Overall, most participants were more efficient when using the integrated representation (*I*) ($\varrho(36) = 0.503$, $p < .05$). This was most apparent for reviews of diagrams with a high degree consistency (*H*) ($\varrho(35) = 0.015$, $p > .05$). For reviews of diagrams with a low degree of consistency (*L*), most participants were equally efficient no matter the representations ($\varrho(36) = 0.414$, $p < .05$).

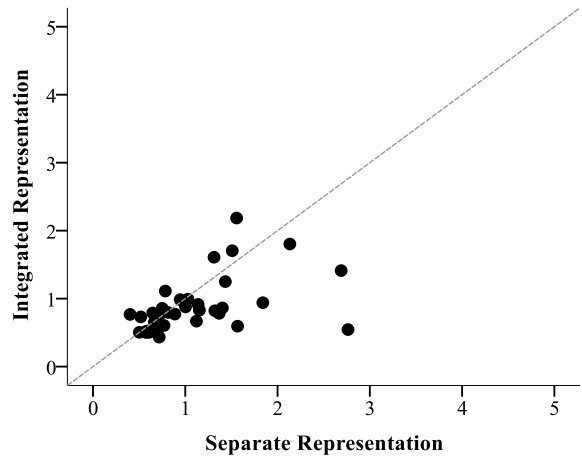

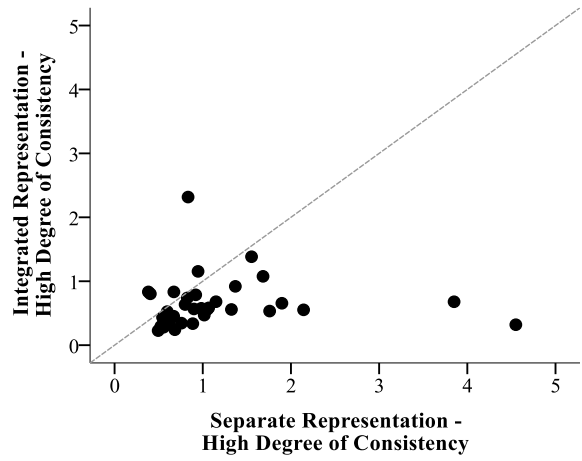

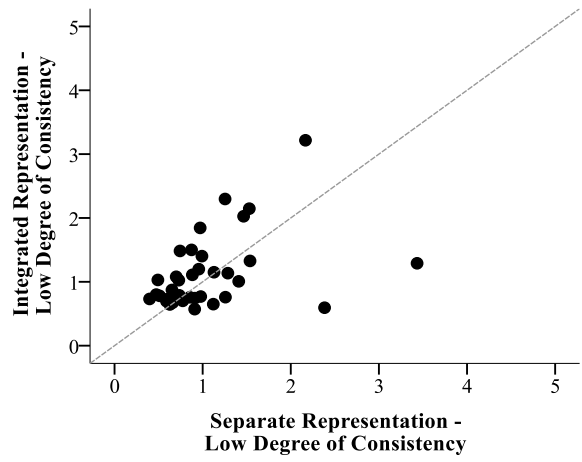

**Fig. 5.** Scatterplots – Efficiency

### 5.3 Hypothesis Tests

This section shows the results of hypothesis tests. As preconditions of parametric test were satisfied, we conducted two-way repeated measures ANOVAs and followed up significant interaction effects using t tests.

*Effectiveness.* The results of a two-way repeated measures ANOVA indicate no significant main effect of the representation. $F(1, 35) = 0.001$, $p > .05$, d = 0.01, Power: .92. There was, however, a highly significant main effect of degree of consistency. $F(1, 35) = 34.86$, $p < .01$, $d = 0.99$, Power: 1. As there was no significant interaction effect $F(1, 35) = 0.46$, $p > .05$, $d = 0.11$, Power: .18 between representation and degree of consistency, the significant main effect of degree of consistency can be interpreted globally. Since effectiveness is higher when reviewing diagrams with a high degree of consistency (*H*) we accept **HEff2-1a**. However, we cannot reject **HEff1-0** and **HEff3-0**.

*Efficiency.* The results of the two-way repeated measures ANOVA show a significant main effect of representation $F(1, 35) = 4.83$, $p < .05$, $d = 0.38$, Power: .97, and a significant main effect of degree of consistency $F(1, 35) = 5.18$, $p < .05$, $d = 0.39$, Power: .98. There was also a highly significant interaction effect between representation and degree of consistency $F(1, 35) = 9.36$, $p < .01$, $d = 0.52$, Power: .99. This indicates that the representation had different effects on the participants' efficiency depending on the degree of consistency between the reviewed diagrams. We therefore accept **HEfy3-1**.

As the interaction diagrams (cf. **Fig. 6**) show a disordinal interaction between representation and degree of consistency, the significant main effects cannot be interpreted globally.

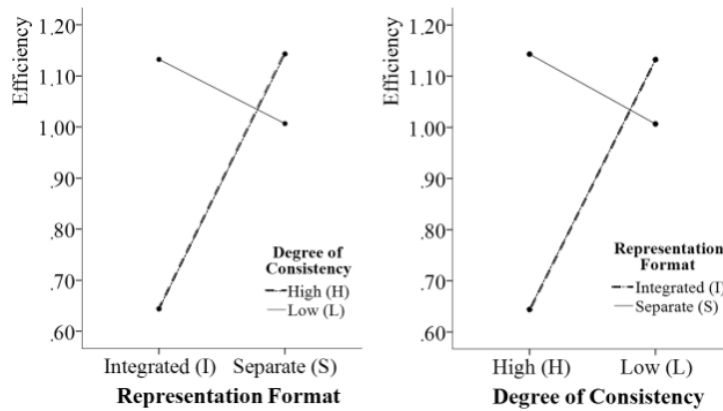

**Fig. 6.** Interaction diagrams – Efficiency

To further investigate the interaction effect, we conducted t tests, where we kept one of the factors constant. Table 5 shows the results for all four combinations. On average, participants reviewed diagrams with a high degree of consistency (*H*) (M = 0.65, σ = 0.39) significantly more efficiently than diagrams with a low degree of consistency (*L*) (M = 1.13, σ = 0.58) when using the integrated representation (*I*) t(34) = -5.30, p < .05, d = 0.89, Power: .99.

**Table 5.** t-test results for efficiency in minutes

|  | Paired Differences | | | | | t | df | Sig.(2-tail) |
|---|---|---|---|---|---|---|---|---|
|  | Mean | Std. Deviation | Std. Error Mean | 95% Confidence Interval | | | | |
|  |  |  |  | Lower | Upper |  |  |  |
| **IH–IL** | -0.488 | 0.546 | 0.092 | -0.676 | -0.301 | -5.293 | 34 | <0.001 |
| **SH–SL** | 0.087 | 0.955 | 0.159 | -0.236 | 0.410 | 0.548 | 35 | 0.587 |
| **IH–SH** | -0.499 | 0.959 | 0.162 | -0.828 | -0.170 | -3.078 | 34 | 0.004 |
| **IL–SL** | 0.073 | 0.638 | 0.106 | -0.144 | 0.289 | 0.681 | 35 | 0.500 |

When using the separate representation (*S*), the participants' efficiency was not significantly higher when reviewing diagrams with a low degree of consistency (*L*) (M = 1.04, σ = 0.60) than when reviewing diagrams with a high degree of consistency (*H*) (M = 1.13, σ = 0.87) t(35) = -0.548, p > .05, d = 0.09, Power: .08.

When reviewing diagrams with a high degree of consistency (*H*), participants were highly significantly more efficient when using the integrated representation (*I*) (M = 0.65, σ = 0.39) than when using the separate representation (*S*) (M = 1.14, σ = 0.88), t(34) = -3.08, p < .01, d = 0.52, Power: .63.

When reviewing diagrams with a low degree of consistency (*L*), participants were not significantly more efficient when using the separate representation (*S*) (M = 1.04, σ = 0.60) than when using the integrated representation (*I*) (M = 1.12, σ = 0.58), t(35) = -0.68, p > .05, d = 0.11, Power: .10.

Note that there are variations in the number of participants, since we could not determine efficiency for participants who made no correct review decision.

## 6   Discussion and Conclusion

### 6.1   Evaluation of Results and Implications

To answer the question whether to review inconsistencies between different stakeholders' intentions on behavioral properties in one integrated diagram or in separate diagrams, the experiment investigated the participants' effectiveness and efficiency for both representation formats.

Regarding *effectiveness*, the representation format itself had no significant impact. When reviewing diagrams with a low degree of consistency the readability of the integrated representation seems to decrease. Additionally, effectiveness is considerably higher when reviewing diagrams with a high degree of consistency than when reviewing diagrams with a low degree of consistency, regardless of the representation.

Regarding *efficiency*, it depends on the degree of consistency whether an integrated or a separate representation is more advantageous when reviewing inconsistent behavioral properties. As the results show, when reviewing diagrams with a high degree of consistency, efficiency is highly significantly higher when using the integrated representation. Since there is no significant difference in effectiveness, we conclude that diagrams with a high degree of consistency can be more efficiently reviewed using the integrated representation without sacrificing accuracy. The fact that there is no statistically significant difference between the representation formats when reviewing diagrams with a low degree of consistency might be due to low statistical power in this case. Surprisingly, participants using the separate representation were more efficient when reviewing diagrams with a low degree of consistency than when reviewing diagrams with a high degree of consistency.

### 6.2   Threats to Validity

To address threats to validity, which exist for this type of study (cf. [43]), we have employed certain mitigation strategies [42]. The most relevant validity threats are discussed below:

*Construct Validity.* The example specification was carefully adopted in close collaboration with industry experts from the avionics domain. In addition, the setup was discussed with industry experts from the automotive domain to ensure for domain independent generalizability. In addition, we used a pretest group to validate the experiment setup and material. To avoid social threats to construct validity such as hypothesis guessing and evaluation apprehension, we did not use extensive briefings, nor did we give bonuses for experiment participation. Since the experiment participants were familiar with reviews and the MSC notation format, we did not use extensive briefing, thus lowering threats from hypothesis guessing by using naïve subjects. Furthermore, we did not give bonuses related to the performance in the experiment or to the participation on the experiment. In addition, we only use quantitative measurements.

However, in case of efficiency, we must discuss some threats to validity arising from the experiment setup. As we used an online questionnaire, which does not measure the time usage for a single decision but for the review of a whole diagram, we can make no statements about the exact time used for reviewing each single stakeholder intention. Since the experiment participation was done online, we have no knowledge about time-consuming activities participants might have done during their experiment participation. In a short briefing, we stressed the need for focused work on the experiment. In addition, we designed the experiment in such a way that the experiment could be completed in less than 30 minutes to minimize the number of participants losing focus. While we removed outliers indicating large irregularities, we cannot eliminate the issue that smaller activities (e.g., chatting or answering phone calls) could have influenced our measurements.

*Internal Validity.* We designed the experiment as an online questionnaire to be conducted within about 30 minutes and gave the participants a time frame of 5 days to participate. Thus, we assume that internal threats to history, maturation, or mortality do not exist. To avoid threats from compensatory equalization, we decided to use a within-subject design. The order of treatments was randomized among all students. In doing so, we avoided single group threats and reduced effects from interactions among subjects. However, it must be noted that allowing a time frame to participate and to allow participation online also relates to losing control over participants' behavior regarding experiment participation. For example, we cannot guarantee that there has been no interaction between participants.

*External Validity.* The participants were mostly graduate students except for a few undergraduates in their senior year. As the participants were students, the question of generalizability to an industrial setting arises. Since studies (cf. [44], [45], [46], [47]) showed that graduate students can serve as an adequate replacement for industry professionals in experiments, and we discussed experiment material, experiment tasks and experiment results with our industry partners, we are confident that the findings can be generalized.

*Conclusion Validity.* To avoid threats from unreliable measures we used expert reviews of the experiment material. In addition, we used a pretest to validate participants' ability of understanding the material in the intended way. Note that pretest participants were not chosen from the set of final participants. Furthermore, we involved industry professionals to ensure that the transformation of industrial examples into experiment material did not corrupt the principle intention of industrial problems and examples.

### 6.3 Inferences

Regarding the question, whether it is beneficial for manual reviews to first merge inconsistent behavioral properties into one integrated diagram compared to the review of inconsistent properties in separate diagrams, the experiment shows that such a model merging seems to have only limited impact on the effectiveness of the review. In contrast to previous work that has shown that the use of model transformations and model merging for consistent behavioral properties does significantly impact effectiveness of the review (cf. [16]), no such overall advantages (or any disadvantages) for inconsistencies were recognizable in this experiment. However, two major findings remain and may provide a starting point for future work:

First, when reviewing models with minor inconsistencies a merging of the inconsistent parts into one diagram can significantly improve the reviews efficiency. Since minor inconsistencies easily occur in model-based engineering (e.g., due to simple misnaming errors), this effect might significantly impact the overall review of an entire specification. Consequently, future work should deal with determining the maximal ratio of inconsistent and consistent parts that should be merged to allow for efficient reviews of the merged diagram.

Second, the results show that regardless of the representation format (i.e. two separate diagrams or one merged diagram) the effectiveness of the review is considerably higher when reviewing diagrams with a high degree of consistency than when reviewing diagrams with a low degree of consistency. Therefore, for manual validation, it might be beneficial to determine the degree of consistency between different views beforehand. Based on the degree of consistency it can then be decided, which representation format should be chosen for the review. Therefore, future work should deal with the question, how best to determine the degree of consistency and what the preferable review format for a certain degree of consistency is. Furthermore, it might also be interesting to investigate the tradeoff between more reviews in shorter time and fewer reviews far between. Hence, the question arises whether shorter reviewing cycles are advantageous to longer review cycles, as inconsistencies are removed earlier, and, consequently, larger inconsistencies are less likely to occur.

**Acknowledgment.** This research was partly funded by the German Federal Ministry of Education and Research (grant no. 01IS16043V and grant no. 01IS12005C). We thank Stefan Beck and Arnaud Boyer (Airbus Defence and Space), Jens Höfflinger (Bosch), and Karsten Albers (inchron) for their support regarding the adoption of industrial specifications to fit as experiment material.